\documentclass{rspublic}
\usepackage{psfig}
\def\[{\begin{equation}}
\def\]{\end{equation}}
\def\lta{\lesssim}
\def\gta{\gtrsim}

\def\kms{{\rm\,km\,s^{-1}}}
\def\msun{\,{\rm M}_\odot}
\def\erg{\,{\rm erg}}\def\ergs{\,{\rm erg\,s}^{-1}}
\def\kB{k_{\rm B}}
\def\fracj#1#2{{\textstyle{#1\over#2}}}
\def\b#1{{\bf#1}}
\def\d{{\rm d}}
\def\K{\,{\rm K}}\def\kpc{\,{\rm kpc}}
\def\yr{\,{\rm yr}}\def\Myr{\,{\rm Myr}}\def\Gyr{\,{\rm Gyr}}
\def\lesssim{\mathrel{\hbox{\rlap{\hbox{\lower4pt\hbox{$\sim$}}}\hbox{$<$}}}}
\def\gtrsim{\mathrel{\hbox{\rlap{\hbox{\lower4pt\hbox{$\sim$}}}\hbox{$>$}}}}
\def\mnras{MNRAS}
\def\apj{ApJ}\def\apjl{ApJ}
\begin{document}

   \title[Black holes, cuspy atmospheres, and galaxy formation]
{Black holes, cuspy atmospheres, and galaxy formation}

   \author[J. Binney]{James Binney}

\affiliation{Theoretical Physics, 1 Keble Road, Oxford OX1 3NP}

\label{firstpage}
\maketitle

\begin{abstract}{Cooling flows -- galaxies: nuclei -- galaxies: formation --
galaxies: jets -- galaxies: luminosity function}    
In cuspy atmospheres, jets driven by supermassive black holes (BHs) offset
radiative cooling. The jets fire episodically, but often enough that the
cuspy atmosphere does not move very far towards a cooling catastrophe in the
intervals of jet inactivity. The ability of energy released on the
sub-parsec scale of the BH to balance cooling on scales of several tens of
kiloparsecs arises through a combination of the temperature sensitivity of
the accretion rate and the way in which the radius of jet disruption varies
with ambient density. Accretion of hot  gas does not
significantly increase BH masses, which are determined by periods of rapid
BH growth and star formation when cold gas is briefly abundant at the
galactic centre.  Hot gas does not accumulate in shallow potential wells. As
the Universe ages, deeper wells form, and eventually hot gas accumulates.
This gas soon prevents the formation of further stars, since jets powered by
the BH prevent it from cooling, and it mops up most cold infalling gas
before many stars can form. Thus BHs set the upper limit to the masses of
galaxies. The formation of low-mass galaxies is inhibited by a combination
of photo-heating and supernova-driven galactic winds.  Working in tandem
these mechanisms can probably explain the profound difference between the
galaxy luminosity function and the mass function of dark halos expected in
the cold dark matter cosmology. 
\end{abstract}

\section{Introduction}

Gravitational potential wells that are deep enough to trap gas hotter than
$\sim3\times10^6\K$ can generally be detected in the thermal X-ray emission
of the trapped gas. These potential wells range in size from those of
massive elliptical galaxies through groups of galaxies to clusters of
galaxies. As one proceeds down this sequence, the fraction of the system's
baryons that are contained in the virial-temperature gas rises from
$\lta10\%$ to $\gta80\%$.

The central cooling time $t_{\rm c}(0)$ is defined to be the ratio of the
central energy density to the central luminosity density due to radiative
cooling.  In many, perhaps most, systems, $t_{\rm c}(0)$ is shorter than the
Hubble time. In the case of an elliptical galaxy such as NGC 4472, $t_{\rm
c}(0)\simeq0.3\Myr$, while in a cluster of galaxies such as Hydra $t_{\rm
c}(0)\simeq300\Myr$. Hence, we must ask how these 
systems endure for times that greatly exceed $t_{\rm c}(0)$.

In the absence of heating, radiative losses cause the central density to
rise inexorably as the central temperature falls. The density reaches
arbitrarily large values in a time $t$ that is slightly shorter than $t_{\rm
c}$ (Murray \& Balbus, 1992). Kaiser \& Binney (2003) present a
semi-analytic model of this process, which ends in a `cooling catastrophe'.

The XMM-Newton and Chandra satellites have established two facts for which
there was sketchy evidence in earlier data.  First, although the temperature
drops as one approaches the centre of one of these systems, it is bounded
below by a `floor' temperature $\sim\fracj13 T_{\rm vir}$, where $T_{\rm
vir}$ is the `virial temperature' characteristic of the bulk of the X-ray
emitting gas. Second, the X-ray emitting plasma is clearly being heated by
outflows from a centrally located active galactic nucleus that is surely an
accreting black hole (BH).  These facts have greatly strengthened the case
that in the long run the energy radiated by the hot gas is replaced by
energy released by accretion of gas onto the BH. Consequently, in these
systems gas is neither cooling nor flowing inwards as has traditionally been
supposed, and their established designation as `cooling flows' is
unfortunate. A more appropriate name is `cuspy atmosphere' since the
defining characteristic of these systems is a sharply peaked X-ray
surface-brightness profile, which proves to be associated with a central
depression in the temperature of the gas. 

Many questions about cuspy atmospheres remain open. These include (1) the
mechanism by which energy is transported from the solar-system scale of the
BH to the $10$ to $1000\kpc$ scale of the thermal plasma, and (2) the
timescale between eruptions of the BH and the corresponding depth of the
excursions in the central density of the cuspy atmosphere.

\section{Time between eruptions}

Two extreme views are possible on this second point. A violent outburst of
the BH might stir the trapped gas into something approaching an adiabatic
atmosphere -- one in which the specific entropy $s$ is everywhere the same.
If heating then stops completely, the specific entropy profile $s(r)$ 
steepens as the system drifts towards a cooling catastrophe, at which
another violent outburst of the BH reheats to a state of near-uniform $s$
(Kaiser \& Binney 2003).  In this picture, systems such as Hydra
and Virgo are observed $\sim300\Myr$ before their next cooling catastrophe.
The opposite extreme was explored by Tabor \& Binney (1993), who conjectured
that steady central heating generates a uniform-entropy core, which
gradually expands as material falls into it at the base of an enveloping
cuspy atmosphere.

Observations cast doubt on this last picture in two ways. First, cuspy
atmospheres appear not to have adiabatic cores (Kaiser \& Binney 2003).
Second, there is much evidence that BHs eject energy in discrete bursts
rather than continuously. 

The absence of adiabatic cores is a clue to the way in which BHs heat the
system. If photons carried the energy from the relativistic region, the
energy would be thermalized deep down and then convected outwards, as it is
in a late-type star with a convective core. If jets carry the energy away
from the BH, it will  thermalize over a wide range of radii, including radii
in excess of the $>100\kpc$ scale of the cuspy atmosphere. So with jet heating
an adiabatic core need not arise (Binney \& Tabor 1995).

The most relevant evidence for discrete bursts of heating also confirms that
jets are the intermediaries: we see `cavities' or `bubbles' in the X-ray
emitting plasma that are surely inflated where a jet is disrupted as it
impacts the denser thermal plasma.  Several systems show more than one
generation of cavity, and the cavities nearer the BH are younger and thus
more luminous in synchrotron radiation from extremely energetic electrons
(Lorentz factors $\gamma\gta10^5$). It is generally agreed that these
cavities move outwards at approximately the speed of sound ($\sim1000\kms$)
in the ambient plasma
(Gull \& Northover 1973; Churazov {\it et al.}\ 2001; Quilis {\it et al.}\ 2001;
Br\"uggen \& Kaiser 2001, 2002; Br\"uggen {\it et al.}\ 2002).

\begin{table}
\caption{Parameters for five clusters with cavities.}
\begin{tabular}{lcccl}
\hline\noalign{\vskip1pt}
system&${\displaystyle\frac{PV}{10^{58}\erg}}$&
${\displaystyle\frac{L_X}{10^{43}\ergs}}$&
${\displaystyle\frac{\tau}{\Myr}}$&reference\\
\noalign{\vskip1pt}\hline
Hydra A&27&30&	88	&{{McNamara {\it et al.} 2000, Nulsen {\it et al.} 2002}}\\
A2052  &4 &3.2&	122	&Blanton {\it et al.} 2001\\
Perseus&8 &27  &29		&Fabian {\it et al.} 2000, Allen {\it et al.} 1992\\
A2597  &3.1&3.8&79	&McNamara {\it et al.} 2001\\
A4059  &22&18&	119	&Huang \& Sarazin 1998, Heinz {\it et al.} 2002\\
\hline
\end{tabular}
\end{table}

Does trapped virial-temperature gas drift far towards a cooling catastrophe
during inter-outburst intervals, so that it must be radically restructured
during an outburst?  The rate of evolution of the density profile
of X-ray emitting gas accelerates slowly at first, and very rapidly towards
the end of a drift to a cooling catastrophe. Hence, if most sources are
drifting towards the next cooling catastrophe, many sources will be seen in
configurations near those produced by an outburst, and only a few sources
will be found close to a cooling catastrophe.  From the fact that $\gta70\%$
of X-ray clusters have cusped cooling cores in which cooling times
$\lta1\Gyr$ occur (Peres {\it et al.}\ 1998), it follows that near-adiabatic
states are not produced by outbursts, and the time between outbursts is
$\lta1\Gyr$. Kaiser \& Binney (2003) concluded that the scarcity of gas at
$\lta\fracj13T_{\rm vir}$ is compatible with the sources cycling between the
least centrally concentrated configurations observed and cooling
catastrophes. 

However, the data do not require such deep cycles. The sizes and locations
of cavities in clusters such as Perseus (Fabian {\it et al.}\ 2000), MKW3s
(Mazzotta {\it et al.}\ 2002) and Abell 2597 (McNamara {\it et al.}\ 2001)
suggest that a new pair of cavities is produced every $\lta50\Myr$, and
simple estimates of the energy injected into the thermal plasma over the
lifetime of a cavity (Churazov et al.\ 2002) suggest that in this case
heating by the BH can balance radiative cooling. Table 1 illustrates this
point by giving for five clusters an estimate of pressure times volume for a
pair of cavities, the X-ray luminosity of the cuspy atmosphere $L_X$ (mostly
from the classic $\dot M$ value) and the characteristic time $\tau=3PV/L_X$.
The {\it minimum\/} work done by an AGN in blowing a cavity is $\fracj52PV$,
rising to $4PV$ if the fluid within the cavity is relativistic.  Since the
inflation of cavities is likely to be highly irreversible, especially in its
early stages and from the perspective of the ambient medium, the actual work
done will be larger.  If we conservatively assume that the work done is
$3PV$, then heating will balance cooling if the intervals between the
creation of cavities equals the quantity $\tau$ listed in table 1. These
intervals agree to within the errors with the values estimated from
hydrodynamic models.

\section{The coupling between BH and X-ray atmosphere}

The suggestion that heating balances cooling is puzzling for two reasons.
First, there is much evidence BHs release energy very
unsteadily, and second because there is a gross mismatch between the $50$ to
$500\Myr$ timescale on which the thermal plasma adjusts its configuration and
the $\lta10\yr$ timescale on which the energy output of massive BHs is known
to change by factors of 2. So what mechanisms could enable a BH to hold
fairly constant the density profile in the vastly bigger atmosphere of X-ray
emitting gas?

\subsection{Bondi accretion}

One condition for establishing a steady state is that the energy output of
the BH should increase when the cuspy atmosphere's central density increases,
and do so with a delay that is small compared to the central cooling time
$\sim100\Myr$. The Bondi radius $r_{\rm Bondi}$ is the distance from the BH
at which the BH's Kepler speed equals the sound speed. In the cases of Sgr
A$^*$ at the centre of the Galaxy and of M87, $r_{\rm Bondi}$, which bounds
the BH's sphere of influence, is resolved by {\it Chandra}. Hence we can be
pretty certain of the rate at which gas flows into the sphere of influence.
Gas with temperature $T_0$ flows at the sound speed $c_s$ through the
spherical surface of area
 \[
A=4\pi\left({GMm_p\alpha\over\kB T_0}\right)^2,
 \]
 where $\alpha=\frac35\mu$, with $\mu=0.62$ the molecular weight.  
The particle density is $n\simeq P/\kB T_0$, where $P$ is the pressure
just above the surface. Hence, the accretion rate
 \[\label{Mdoteq}
\dot M\simeq n\mu m_pAc_s\simeq7
\pi P(GM)^2\left({\alpha m_p\over\kB T_0}\right)^{5/2}
\]
 rises at least as fast as $T_0^{-5/2}$, and any drop in
$T_0$ will quickly lead to an increase in the BH's power output. The
luminosity of the cooling core is
 \[\label{Leq}
L=\int\d^3\b x\,n^2\Lambda(T)\simeq\int\d^3\b x\,\left({P\over\kB
T}\right)^2\Lambda ,
\]
 which has a very similar dependence on $T_0$ when one takes into account
the tendency of $P$ to rise slightly as $T_0$ declines. Hence the mass
falling into the sphere of influence during an outburst is expected to be
roughly proportional to the energy radiated by the thermal plasma in the
cooling core.

We are unsure what fraction of the material that enters the sphere of
influence is accreted by the BH rather than being blown out in a wind or
jet. How much energy is released when a given mass of gas is swallowed by
the BH is also controversial. Fortunately observations of the best observed
system suggest resolutions of these questions.

In M87 the Bondi accretion rate (\ref{Mdoteq}) is $0.1\msun\yr^{-1}$, which yields a
luminosity $5\times10^{44}\ergs$ if $0.1mc^2$ of energy is released for
accretion of mass $m$ onto the BH (Di Matteo {\it et al.} 2003). The X-ray
luminosity from the central $\sim20\kpc$ of the cuspy atmosphere is
$10^{43}\ergs$ (Nulsen \& B\"ohringer 1995), while that of the AGN is
$<5\times10^{40}\ergs$. Estimates of the mechanical luminosity of the jet
that emerges from the AGN range from $10^{43}\ergs$ (Reynolds et al 1996) to
$10^{44}\ergs$ (Bicknell \& Begelman 1999; Owen et al 2000). Thus the data
for M87 suggest that the BH is accreting at a substantial fraction of the
Bondi rate, and that the energy released is passing along the jets to reheat
the cuspy atmosphere on $10\kpc$-scales. Radiative losses near the BH are
negligible. This is precisely the situation envisaged by Binney \& Tabor (1995).

Material that falls into the sphere of influence is likely to form an
accretion disk or torus. In the case of M87, a disk of ionized gas has
actually been seen with HST (Harms {\it et al.} 1994). The accretion disk
will introduce a lag between a drop in the central temperature of the
cuspy atmosphere and an increase in the power of the BH equal to the time it
takes material to spiral through the disk.  If this delay exceeded
$\sim50\Myr$, large-amplitude feedback oscillations would probably occur.
Evidence that cuspy atmospheres are in near steady-states therefore suggests
that material either accretes directly from the Bondi flow, or spirals
through the accretion disk in $\lta50\Myr$.

\subsection{Fixing the radial density profile}

For a steady state to be reached, the radial profile of energy deposition by
the jets at outburst must coincide with the radial profile of radiative
losses between outbursts. Energy is transferred from a jet to the ambient
plasma when the latter disrupts the jet, either in part or totally, as at the
hot spot of a Fanaroff--Riley (FR) II radio source. The more powerful a jet is, the further
out it will go before it is strongly disrupted. So the fraction of a jet's
energy that is deposited at large radii should increase with jet power.
Conversely, the higher the ambient density is near the BH, the
smaller will be the radii at which a given jet is disrupted and its energy
thermalized. 

Motivated by these considerations, Omma \& Binney (2004) repeatedly
simulated the dynamical evolution of cluster gas from an initial state that
resembles the current state of the Hydra cluster. Each simulation was fully
three-dimensional and used the adaptive-mesh code ENZO (Bryan 1999). 
The rate of radiative cooling was calculated for an optically thin plasma in
thermal equilibrium. Hence a cooling catastrophe arose in the absence of jet
heating.

In simulation 1 the jets fire after $262\Myr$ of cooling. They have a total
power of $5\times10^{44}\ergs$ and run for $25\Myr$, during which time they
inject $4\times10^{59}\erg$. The jets in simulation 2 fire after $300\Myr$
of cooling, by which time an extra $4\times10^{59}\erg$ has been lost to
radiation, and they inject $8\times10^{59}\erg$ at $10^{45}\ergs$. Thus the later ignition of
the jets in simulation 2 is compensated for by enhanced energy injection
along the lines suggested by the model of Bondi accretion.  

It is instructive to compare these energies with what would be available
through Bondi accretion onto a BH of mass $M$ under the assumption that
accretion of mass $m$ by the BH releases $0.1mc^2$ of energy. If the
atmosphere were isothermal in the numerically unresolved region from a
radius $1\kpc$ to the BH's radius of influence, the energy available from
Bondi accretion would be $E=5(M/10^9\msun)^2\times10^{59}\erg$ over
$262\Myr$, and $7(M/10^9\msun)^2\times10^{59}\erg$ over $300\Myr$. Thus for
black hole masses $\sim3\times10^9\msun$ of the expected order, $\sim10\%$
of what flows into the BH's sphere of influence needs ultimately to be
accreted by the BH. For a BH of this mass the BH's mechanical luminosity in
simulation 1 is $7\times10^{-4}L_{\rm Edd}$, where $L_{\rm Edd}$ is the
Eddington luminosity at which free-electron scattering causes radiation
pressure to balance gravity.

\begin{figure}
\centerline{\psfig{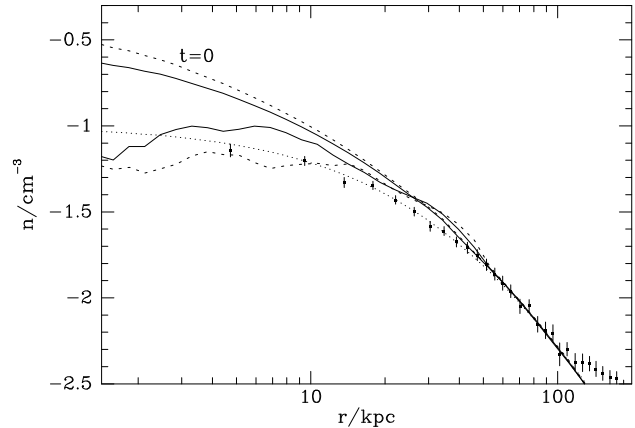}}%
\caption{Dotted curve: initial density profile of all simulations. Data
points: density in Hydra from David {\it et al.}\ (2000).  Curves
labelled $t=0$: densities
after cooling and immediately before jet ignition in simulation 1 (full
curve) and simulation 2 (dashed curve). Bottom  curves show
spherically averaged density profiles $42\Myr$
after ignition in simulation 1 (full) and simulation 2 (dashed).
\label{densfig}}
\end{figure}

In figure \ref{densfig} the dotted curve shows the density profile of the
cluster gas at the start of both simulations. The data points show the
density in the Hydra cluster as deduced by David {\it et al.}\ (2000). The
upper full
curve shows the density profile at the ignition of the jets
in simulation 1, while the dashed curve labelled $t=0$ shows the
density profile at the ignition of the jets in simulation 2. The
effect on the density profile of $\sim300\Myr$ of passive cooling is
evident.  The bottom full curve shows the spherically averaged density
profile $42\Myr$ after the firing of the jet in simulation 1, while the
bottom dashed curve shows the same data for simulation 2. At that time,
$17\Myr$ after the jets extinguished, the curves are quite similar to the
initial profile and the data. Thus in both simulations the injected energy
has effectively reversed the effect of $300\Myr$ of cooling. 

Most crucially, the dashed curve of simulation 2 now lies {\it below\/} the
full curve, implying that the system that cools for longer and has the most
centrally concentrated density profile when its jets ignite, ends up with
the {\it less\/} centrally concentrated profile.  The density profiles at
times later than those shown in figure~\ref{densfig} confirm that the greater
central concentration of simulation 1 at $t=42\Myr$ is not an aberration:
the profile for simulation 1 remains on top of that of simulation 2, and
moves upwards faster. Consequently, when the profiles are next similar to
those labelled $t=0$ in figure~\ref{densfig}, we can expect simulation 1 to be
the scene of the more energetic outburst slamming into the more centrally
concentrated ICM. When the dust settles after this second outburst, the
profile of simulation 2 will  be the more centrally concentrated and
the pair of simulations will have come full cycle.  Hence these simulations
suggest that the density profiles of cuspy-atmosphere
clusters are oscillating around an attracting profile.

\section{Impact on galaxy formation}

It has long been suggested that the cuspy-atmosphere phenomenon is fundamental
for the galaxy-formation process (e.g., Fabian 1994). I agree,
but I want to persuade you that cuspy atmospheres do not tell us how galaxies
formed, but why they ceased forming (Binney 2004).

The standard picture of galaxy formation starts from the assumption that
when gas falls into a potential well, it shock heats to the virial
temperature (Rees \& Ostriker 1977; White \& Rees 1978). There is increasing
evidence that this assumption is significantly misleading: only a fraction
of infalling gas is heated to the virial temperature, and this fraction is
large only for potential wells that are deeper than those associated with
galaxies (Binney 1977; Katz {\it et al.}\ 2003; Birnboim \& Dekel, 2003). On
account of the shape of the cooling curve of optically thin plasma, the
temperature of infallen gas is bimodal. It seems likely that stars form from
a fraction of the cold gas, and energy released by these stars strongly
heats the remaining gas. If the potential well has a virial velocity below
$\sim 100\kms$ (roughly that of an $L_*$ galaxy), the heated gas flows out
of the potential well and star formation ceases until more cold gas can fall
in. Through repeated accretion of cold gas, a disk galaxy slowly builds up.
A merger may convert this to an early-type galaxy, but subsequent infall of
cold gas and star formation can restore its status as a disk galaxy.

Through mergers and gas accretion, the depth of the potential well
increases.  When its virial velocity reaches $\sim 100\kms$, gas heated by
star formation can no longer be driven out (Dekel \& Silk, 1986).
Consequently, an atmosphere of virial-temperature gas builds up.  Such
atmospheres have been called a cooling flows because their central cooling
times are short. Actually the temperature of such a system is
thermostatically controlled by the nuclear BH, which grew to its current
size during merging episodes as large quantities of cold gas were driven to
the centre, stimulating bursts of star formation, and permitting the BH to
gorge itself at $L_{\rm Edd}$.

As the density and temperature of the virial-temperature atmosphere
increases, the environment becomes hostile to cold gas: filaments of
infalling cold gas are  shredded by Kelvin-Helmholtz
instability and evaporated by electron conduction (Nipoti \& Binney, 2004).
This evaporation of cold gas can happen far from the BH, although the energy
required to heat the cold gas ultimately comes from the BH, which
underwrites the atmosphere's temperature. The elimination of filaments of
infalling cold gas gradually throttles star formation, because the hot
atmosphere never produces cold gas: the coldest part of the atmosphere
surrounds the BH, and energy released by the BH reheats it long before it
can reach the kinds of temperatures ($\lta30\K $) at which stars can form.
The effect of the hot atmosphere on the star-formation rate is not sudden,
however, because a sufficiently massive filament on a sufficiently
low-angular-momentum orbit can always get through to the atmosphere's
cooling core, where it can survive thermal evaporation for a significant
time and lead to the formation of some stars. In the centres of clusters
such as Perseus we see such filaments and infer that they have embedded star
formation (McNamara {\it et al.}\ 1996; Conselice {\it et al.}\ 2001; Fabian
{\it et al.}\ 2003). These filaments have often been supposed to
have formed through catastrophic cooling of the hot atmosphere, but their
dust content and morphology are more consistent with the infall hypothesis
(Soker {\it et al.}\ 1991; Sparks {\it et al.}\ 1989; Sparks 1992).

The galaxy luminosity function differs profoundly from the mass function of
dark-matter halos in all cold-dark matter (CDM) cosmogonies. Specifically,
there are both fewer faint galaxies than low-mass halos, and fewer luminous
galaxies than high-mass halos. The dearth of low-luminosity galaxies can be
plausibly ascribed to the effects of photoionization at redshifts $z\lta20$
(Efstathiou 1992, Dekel 2004) and to the ability mentioned above, of star formation
to heat residual gas and drive it out of shallow potential wells. In a
recent examination of this problem in the context of semi-analytic galaxy
formation models, Benson {\it et al.}\ (2003) found that when feedback was strong
enough to make the number of low-luminosity galaxies agree with observation,
too many high-luminosity objects formed because gas ejected from shallow
potential wells later fell into deep potential wells. The crucial ingredients
missing from the Benson {\it et al.}\ models are (a) the ability of the central
black hole to prevent cooling of virial-temperature gas, and (b) the infall
of cold gas,
together with the tendency of a hot atmosphere to destroy filaments of
cold infalling gas.

\subsection{History of BH growth}

The demography of quasars and radio galaxies indicates that most of the
energy released in the formation of a massive BH has emerged in bursts of
accretion that have driven the luminosity to near $L_{\rm Edd}$. Thus Yu \&
Tremaine (2002) found that the total energy emitted in the optical and UV
bands by AGN lies remarkably close to the energy released in the growth of
massive BHs. They also showed that a high efficiency $\epsilon\gta0.1$ for
the conversion of accretion energy to optical/UV photons and radiation at
$L\simeq L_{\rm Edd}$ must be assumed if the formation of the known
population of BHs is to generate as many luminous quasars as are observed.
The mass of a BH that radiates at $L_{\rm Edd}$ exponentiates on the
Salpeter time $t_{\rm S}\simeq2.5\times10^7(0.1/\epsilon)\yr$. If BHs form
with masses $M\sim10^3\msun$, then they require $\sim14t_{\rm S}$ to grow to
their current $M\sim10^9\msun$. Thus Yu \& Tremaine require them to have
radiated at $\sim L_{\rm Edd}$ for $\sim0.4\Gyr$ and accreted at $\lta0.05
L_{\rm Edd}$ for the remaining $13\Gyr$.

The tight correlation between BH mass and the velocity dispersion of the
host spheroid tells us that BH growth is dominated by periods of rapid
formation of spheroid stars. This conclusion is reinforced by the similarity
in time and space of the densities of luminous quasars and luminous
star-forming galaxies.  It seems clear that these episodes of rapid BH
and spheroid growth occur when there is plenty of cool gas at the galaxy
centre.  These episodes are short because a combination of star formation,
aided by radiation from the BH ({\sc Ostriker}), and mass loss in a galactic wind,
quickly lowers the gas density to the point at which it can be heated to
$\sim3\times10^6\K$.  Star formation and BH growth then all but cease.  The
hot gas flows out of shallower potential wells, but is confined by wells
with virial velocities $\gta 100\kms$.

Once the host potential is deep enough to trap supernova-heated gas, and a
hot atmosphere builds up, the BH becomes more regularly active.  Its mode of
operation changes significantly, in the sense that its energy output
becomes predominantly mechanical. In general terms it is natural that
photons should diminish in prominence as distributors of the BH's energy
production once the BH starts accreting optically thin, virial-temperature
gas. But  this mode switch has yet to be properly understood.
Observations of M87 leave no doubt that the switch occurs, however. 

The rate at which the BH's mass grows in the new regime is determined by the
rate at which the cuspy atmosphere radiates, which for a typical cluster lies in
the range $10^{43}$ to $10^{44}\ergs$. At the canonical $10\%$ accretion
efficiency, these luminosities imply mass accretion rates $\dot M\sim0.02$
to $0.002\msun\yr^{-1}$. Growth at rates of this order for $10^{10}\yr$ does
not have a significant impact on the mass of a BH that already contains
$>10^9\msun$. This is why the combined radiative output of quasars already
accounts for the observed BH mass density.

Fresh supplies of cold gas can revive star formation. If the gas has high
angular momentum and accumulates in a disk, there can be significant star
formation without enhanced BH growth. Gas infall enhances BH growth only if
the gas tumbles to the galactic centre to form spheroid stars. Hence the BH's
mass is correlated with the properties of the spheroid rather than with
those of the whole galaxy.

\section{Conclusions}

Although some loose ends remain, I am impressed by the way in which
disparate strands of theory and observation are coming together to form a
coherent picture of galaxy formation, and the symbiosis of BHs with
spheroids and cuspy atmospheres. In this picture BHs play a major role in
preventing gas that is heated to the virial temperature from cooling.
Galaxies that are in potential wells deep enough to trap supernova-heated
gas soon cease to form stars, even from cold infalling gas of high angular
momentum, because trapped gas at the virial temperature evaporates infalling
gas before stars can form from it. Hence BHs cause the cutoff above $L_*$ in
the galaxy luminosity function.

BHs are effective thermostatic heaters for two reasons. First, they sample
the coldest gas, and the rate at which they are fed increases rapidly as the
temperature of this gas falls. Moreover, the mass accreted in any time
interval is roughly proportional to the energy radiated by the central part
of the atmosphere in that interval. Second, they inject energy through
jets, and the radial range over which a jet's energy is thermalized
is smaller when the pre-outburst atmosphere is more centrally concentrated.
This phenomenon causes the density profile of an atmosphere to fluctuate
around an attracting profile that appears to be similar to those observed.

For more than two decades the theory of steady-state cooling flows with mass
dropout held up progress in appreciating the role that AGN play in
structuring galaxies. This theory was finally swept away by damning evidence
from the {\it Chandra\/} and {\it XMM-Newton\/} missions, but its internal
contradictions should have left it without proponents a decade ago.
Meanwhile, evidence has emerged for the intimate connection between BHs and
both quasars and spheroids, and for the strength of mass outflows from
star-forming galaxies. The last piece of the jigsaw that gives a coherent
picture of galaxy formation is the still tentative evidence for the
importance of cold infall.

Over the next decade I hope that this picture will be consolidated by
understanding (a) the relation between extra-planar gas around spiral
galaxies and the infall and outflow phenomena, and (b) why BH accretion
sometimes produces the Eddington luminosity and rapid BH growth, and in
other circumstances yields jets with high efficiency. Finally, we must come
to an understanding of how it is that infalling cold gas frequently has enough
angular momentum to form a galactic disk.

\label{lastpage}
\end{document}